\documentclass[12pt,epsf]{article}
\usepackage{bm}
\usepackage{graphicx,color}
\begin{document}
%%%%%%%%%%%%%%%%%%%%%%%%%%%%%%%%%%%%%%%%%%%

%\def\beq{\begin{eqnarray}}
%\def\eeq{\end{eqnarray}}
%\newcommand{\gsim}{ \mathop{}_{\textstyle \sim}^{\textstyle >} }
%\newcommand{\lsim}{ \mathop{}_{\textstyle \sim}^{\textstyle <} }

%%%%%%%%%%%%%%%%%%%%%%%%%%%%%%%%%%%%%%%%%%%%%%%%%%%%%%%%%

\baselineskip 0.7cm

\begin{titlepage}

\begin{flushright}
IPMU16-0128\\
HUPD1607
\end{flushright}

\vskip 1.35cm
\begin{center}
{\large \bf
CP violating phase from minimal texture neutrino mass matrix:

Test of the phase relevant to leptogenesis
}
\vskip 1.2cm
Masataka Fukugita$^{a,b}$,%\footnote{E-mail address: fukugita@icrr.u-tokyo.ac.jp},~ 
~Yuya~Kaneta$^{c}$,%\footnote{E-mail address: kaneta@muse.sc.niigata-u.ac.jp}
~Yusuke Shimizu$^d$,\\%\footnote{E-mail address: yu-shimizu@hiroshima-u.ac.jp}, \\
Morimitsu Tanimoto$^e$%\footnote{E-mail address: tanimoto@muse.sc.niigata-u.ac.jp},~ 
~and~  Tsutomu T. Yanagida$^a$
\vskip 0.4cm
$^a${\it \normalsize 
Kavli Institute for the Physics and Mathematics of the Universe, \\ University of Tokyo, Kashiwa 277-8583, Japan \\}
$^b${\it \normalsize 
Institute for Advanced Study, Princeton, NJ08540, U. S. A. \\}
$^c${\it \normalsize
Graduate~School~of~Science~and~Technology,~Niigata University, \\
Niigata~950-2181,~Japan \\} 
$^d${\it \normalsize
Graduate School of Science, Hiroshima University,  \\
 Higashi-Hiroshima, 739-8526, Japan \\}
$^e${\it \normalsize 
Department of Physics, Niigata University, Niigata 950-2181, Japan \\}

\vskip 1.5cm

\abstract{
The model of neutrino mass matrix with minimal texture is now
tightly constrained by experiment so that it can yield a
prediction for the phase of CP violation. This phase is predicted
to lie in the range $\delta_{CP}=0.77\pi - 1.24\pi$. If neutrino
oscillation experiment would find
the CP violation phase outside this range, this means that
the minimal-texture neutrino mass matrix, the element of which is all real, 
fails and the neutrino mass matrix must be complex, 
i.e., the phase must be present that is responsible for leptogenesis.}

\end{center}
\end{titlepage}

\setcounter{page}{2}

%\section{Introduction}

Following the discovery of neutrino oscillation, we proposed a
neutrino mass matrix with the minimal texture (hereafter FTY model)
\cite{FTY93} assuming that neutrinos are of the Majorana type, to
understand a very large mixing between $\nu_\mu$ and $\nu_\tau$ found
in atmospheric neutrinos.  The 3$\times$3 Dirac neutrino mass matrix
has off-diagonal (1,2) and (2,3) elements in addition to one diagonal
(3,3) element \cite{Fritzsch}. We assumed 3 degenerate right-handed
neutrino masses for economy.  By virtue of the seesaw mechanism
\cite{Minkowski,GRSY} the mixing angle is a quartic root, rather than
a square root \cite{weinberg}, of the neutrino mass ratio and hence it
can readily be large. This matrix was shown to give
empirically determined mixing angles at a good accuracy
\cite{Fukugita:2012jr}.  There appeared much information as to
neutrino mixing over the last two decades, but this matrix so far passed all
critical passes.  Notably, it predicted a finite mixing angle
$\theta_{13}$, which was established by now
\cite{An:2015rpe,Kim:2016yvm,Abe:2014bwa}, and the exclusion of 
maximal mixing of $\theta_{23}$ \cite{NOVA}.

Modern knowledge of the mixing angles allows an accurate and tight
determination of the matrix element.  The final, yet-to-be-known is
the phase of CP violation, $\delta_{CP}$. 
%%%%%%%%%%%%%%%%%%%%%%%%
\footnote{For an earlier attempt, see  ref.\cite{Zhou:2004wz}.}
%%%%%%%%%%%%%%%%%%%%%%%%
  A finite CP phase is being
indicated in recent neutrino oscillation experiments \cite{T2K0,Adamson:2016tbq}.
The prime interest in this phase may be its possible role in
leptogenesis \cite{Fukugita:1986hr}. We must first emphasise, however,
that the phase that is visible in neutrino oscillation is not the
phase that controls leptogenesis: $\delta_{CP}$ being finite does {\it
  not} mean leptogenesis.  The phase that appears in neutrino
oscillation arises from both neutrino and charged lepton sectors. A
finite $\delta_{CP}$ may appear even if the neutrino mass matrix is
real, and this is the case with the original FTY model.
  This, on the
other hand, gives rise to the idea for an important test for the phase
relevant to leptogenesis: whether the experimentally observed phase in
neutrino oscillation deviates from the phase that is predicted from
real matrices is a decisive test for a phase needed for leptogenesis.

%%%%%%%%%%%%%%%%%%%%%%%%%%%%%%%%%%%%%%%%%%%%%%%%%%
  
In this paper, we predict the CP violating phase in our minimal 
texture of the neutrino mass matrix
in  light of new data of T2K \cite{T2K} and NO$\nu$A \cite{NOVA} experiments.
We show that $\delta_{CP}$ is narrowly constrained with the presently
available mixing data.

Our model \cite{FTY93,FTY03,Fukugita:2012jr},  in the basis where
the right-handed Majorana mass matrix is real diagonal, consists of 
mass matrices for the charged lepton and for the Dirac neutrino
of the form \cite{Fritzsch},
\begin{eqnarray}
m_\ell = \left( \matrix{0 & A_\ell & 0 \cr A_\ell & 0 & B_\ell \cr
                     0 & B_\ell & C_\ell\cr        } \right)\ ,\qquad\quad
m_{\nu D} = \left( \matrix{0 & A_\nu & 0 \cr A_\nu & 0 & B_\nu \cr
                     0 & B_\nu & C_\nu \cr        } \right )\  , 
\end{eqnarray}
where each entry is complex in general: in our convention right-handed
fermions operate from the left and left-handed from the right of the
matrix. The phase of $\nu_R$ is fixed by this convention.
We have five phases in the neutrino mass matrix, but three of
them can be absorbed into the wave functions of left-handed doublets
and two were left, say those of $B_\nu$ and of $ C_\nu$ in the third row. 
%%%%%%%%%%%%%%%% add %%%% 
%Note that these phases, in principle, include
%  those characteristic of Majorana neutrinos.
%%%%%%%%%%%% Add %%%%%%%%
 In the FTY model we have assumed these two to be real to
  make the problem analytically tractable, i.e., the elements of the
  neutrino mass matrix $M_R$ and $m_{\nu D}$ are all real. 
The result
determined thereof turns out to agree with neutrino experiment accurately
\cite{Fukugita:2012jr} within the currently available accuracy.
We retain this reality assumption in order to
keep CP invariance in the neutrino sector.  For the charged lepton
mass matrix, three among the five phases are absorbed into the wave
function of right-handed charged leptons, and hence two are left to
us.  This reality of the neutrino mass matrix is pivotal
  in the argument developped in this paper.

We start our analysis with the unit matrix for the right-handed
Majorana neutrino as in the FTY model, 
\begin{eqnarray}
M_R = M_0 {\bf 1},
\label{Majorana}
\end{eqnarray}
i.e., $M_{R1}=M_{R2}=M_{R3}$, but this assumption is {\it ad~hoc} and
will be relaxed to accommodate possible hierarchy in $M_R$.  In this
case the difference in masses (eigenvalues) can be absorbed into
the wave functions of the right-handed neutrinos, which in turn leads
to the violation of the symmetric (i.e., minimal) matrix structure of
the Dirac neutrino mass.  We later relax the assumption of degenerate
$M_{R}$ and introduce parameters,
\begin{equation}
K_{31}=M_{R3}/M_{R1}~,~~~~~~~~~~~~ K_{32}=M_{R3}/M_{R2}~,
\end{equation}
to extend our model, avoiding the {\it ad~hoc} assumption,
but keeping reality of  the matrix elements.
We still take a diagonal basis for $M_R$.
%%%%%%%%%%%%%%%%%%%%
\footnote{See ref.\cite{Obara:2006ny} for an attempt in a similar direction.}
%%%%%%%%%%%%%%%%%%%%%

We remark that the neutrino mass is stable against radiative
corrections.  For the heavy right-handed neutrino of mass
$O(10^{10})\ {\rm GeV}$ the Yukawa coupling for the Dirac neutrino
mass is smaller than $10^{-2.5}$.  A calculation with the
normalization group equation (e.g., \cite{Hagedorn:2004ba}) 
gives the radiative correction of the order $10^{-6}$ relative 
to the leading term, which is negligible.

We obtain the three light neutrino masses, $m_i$ ($i=1,2,3$), as
\begin{equation}
m_i=\left (U_\nu ^Tm_{\nu D}^TM_R^{-1}m_{\nu D}U_\nu \right )_i.
\end{equation}
With the real Dirac neutrino mass matrix the lepton mixing matrix is given by
\begin{eqnarray}
 U = U_\ell^\dagger \ Q \ U_\nu,
\end{eqnarray}
where the expressions of $U_\ell$ and $\ U_\nu$, all their elements
being real, are explicitly given in \cite{FTY03} in terms of the
charged lepton mass and the neutrino mass, and $Q$ is a phase matrix,
\begin{eqnarray}
Q= \left(\matrix{ 1 & 0 & 0 \cr 0 & e^{i \sigma} & 0\cr  0 & 0 & e^{i \tau}
\cr } \right)\ ,
\end{eqnarray}
which corresponds to the two phases left in the charged lepton mass
matrix.
%\footnote{Even if there are phases in the left-handed sector of the
%  Dirac neutrino mass matrix, those can be removed by the phase redefinition.}
Then, the CP violation in neutrino 
oscillations is
written through phases $\sigma$ and $\tau$.
For $U$ we take the conventional $3\times3$ representation used by
Particle Data Group.
\footnote{ In case of the Majorana neutrinos unitary neutrino mixing matrix $U_{\nu}$ may be cast into the form $U_\nu=VP$, where $P$
  is a diagonal phase matrix with two Majorana phases. In our parameterization
  these phases are transferred into $\sigma$ and $\tau$.}

With the charged lepton masses, $m_e$, $m_\mu$, $m_\tau$, given, the
number of parameters in our model is six, $m_{1D},\ m_{2D}, \ m_{3D}$,
$\sigma$, $\tau$ and $M_0$, where $M_0$ is basically fixed by the
neutrino mass, so that the number of parameters is
five that are to be determined from empirical neutrino mixing angles.
%The Majorana phases are included in the left-handed Majorana neutrino masses. 
 We note that this is the minimum texture of the $3\times3$
neutrino mass matrix, in the sense that reducing one more matrix
element (i.e., letting A, B or C to zero) leads to the neutrino mixing
that is in a gross disagreement with experiment.  An antisymmetric
mass matrix with 3 finite elements also leads to a gross disagreement,
as one can readily see. So, our matrix is essentially necessary and sufficient, 
the unique form of
the neutrino mass matrix under the requirement of minimum structure, i.e.,
four texture zeroes.

The lepton mixing matrix elements can be analytically computed, and
are given approximately by the expression as written in Eq.(6) of Ref.\cite{Fukugita:2012jr}.
%\footnote{ We note that the set of
%  equation is a good approximation when normal-hierarchical neutrino
%  masses. For other cases it still gives a correct trend.}.
It has been shown \cite{FTY03} that only normal neutrino mass hierarchy 
is allowed: the model does not accommodate inverted hierarchy nor
%(as seen from the approximate expression of $|U_{\mu 3}|$
%which should empirically be around $1/\sqrt{2}$).  
%Our matrix does not allow 
degenerate neutrinos. It is also shown that $|U_{e3}|$ cannot be too small.
%which require $|U_{e2}|$ close to $1/\sqrt{2}$, while we have
%$|U_{e3}| \approx %|U_{\mu 3}|^2|U_{e2}U_{\mu3}|=
%|U_{\mu 3}|^3|U_{e2}|\sim 0.2$.

We obtain neutrino mass matrix elements, including the phase
$\tau$ and $\sigma$,
with Monte Carlo sampling in 5 parameter space.
We adopt the data \cite{Gonzalez-Garcia:2015qrr}, taking $2\sigma$ as the limit:
\begin{eqnarray}
&&\Delta m_{23}^2= (2.457 \pm 0.047) \times 10^{-3}{\rm eV}^2 \ , \qquad
\Delta m_{12}^2= 7.50^{+0.19}_{-0.17} \times 10^{-5}{\rm eV}^2 \ , \nonumber \\
&&\nonumber \\
&&\sin^2\theta_{12}=0.304^{+0.013}_{-0.012}\ , \quad 
\sin^2\theta_{23}=0.452^{+0.052}_{-0.028}\ , \quad  \sin^2\theta_{13}=0.0218\pm 0.0010 \ ,
\label{data}
\end{eqnarray}
where %we assume the normal mass hierarchy of neutrinos. 
$\Delta m_{23}^2$ and $\Delta m_{12}^2$ represent mass difference
squares relevant to atmospheric neutrino and solar neutrino experiments. 
%%%%%%%%%%%%%%%%%%%%%%%%%%% 
We remark that all oscillation data are fit with
our model well within one sigma errors of experiment.
%%%%%%%%%%%%%%%%%%%%%%%%%%%%%%%%
We take the lowest neutrino mass $m_1$ as a free parameter, while  $m_2$
and $m_3$ are fixed by the mass differences.

%%%%%%%%%%%%%%%%%%%%%%%%%%%%%%%%%%%%%%%%%%%%%%%%%%%%%%%%%%%%%
%%%%%%%%%%%%%%%%%%      Prediction of Mixing Angles    %%%%%%%%%%%%%
%%%%%%%%%%%%%%%%%%%%%%%%%%%%%%%%%%%%%%%%%%%%%%%%%%%%%%%%%%%%%

All neutrino mass parameters are specified in our model, 
as given in Table 1, where
errors shown are at 2$\sigma$. We also added the prediction for
$\sin^2\theta_{23}$, the experimental information for which still has
a large errors and is not restrictive to the model.
In the second column we assume $K_{31}=K_{32}=1$,
the original FTY model.
We predict the effective mass that appears in neutrinoless double 
beta decay 
\begin{eqnarray}
m_{ee}=\left |\sum _{i=1}^3m_iU_{ei}^2 \right|  ~ ,
\end{eqnarray}
to lay in the range $m_{ee}=3.6-5.0$ meV.
The phase parameters $\sigma$ and $\tau$ are well constrained to lie around
$(\pi/2, 3\pi/2)$ or $(3\pi/2, \pi/2)$, where the former range
disappears if $\delta_{\rm CP}>\pi$. This will give the phase derived in
neutrino oscillation, even if the neutrino mass matrix elements are all real.

We note that $\sin^2 \theta_{23}$ is constrained to lie in the range
$0.40-0.47$, where the upper limit comes from empirical $\theta_{13}$, which 
is restricted to a narrow range by modern experiments.  This means that maximal
mixing, $\sin^2 \theta_{23}$=1/2, is not allowed for $\theta_{23}$,
which agrees with the recent experiment \cite{NOVA}.

%%%%%%%%%%%%%%%%%%%%%%%%%%%%%%%%%%%%%%%%%%%%%%

\begin{table}
\caption{parameters of the neutrino mass matrix. The errors stand for
2$\sigma$.}
\begin{center}
\begin{tabular}{lll}
\hline
 parameters      & ~$K_{31}=K_{32}=1~~~$  & ~~~~~$K_{31} \ne 1$ \& $K_{32}\ne\ 1$  \cr
\hline
 && \\
 $ m_1({\rm meV})$ & $1.2^{+0.6}_{-0.5}$ & ~~~~$0-6.6$ \cr 
 $m_2({\rm meV})$  & $8.7^{+0.3}_{-0.2}$ &~~~ $8.7^{+2.3}_{-0.3}$ \cr 
 $ m_3({\rm meV})$ & $49.5\pm 1.0$& ~~~ $49.5\pm 1.5$\cr
 $\sum m_i({\rm meV})$ & $59.5\pm 1.5$&~~~~$58.5^{+9.3}_{-1.5}$ \cr
 $ m_{ee}({\rm meV})$ & $4.2^{+0.8}_{-0.6}$ &~~~~$3.6^{+5.2}_{-0.5}$ \cr
 $\sin^2\theta_{23}$ & $0.45^{+0.02}_{-0.05}$ & ~~~~$0.46^{+0.10}_{-0.06}$ \cr  
 $\delta_{CP}({\rm radian})$ & $(0.89^{+0.11}_{-0.12})\pi$, \ $(1.11^{+0.13}_{-0.11})\pi$&
~~~ $(0.83^{+0.17}_{-0.09})\pi$, \ $(1.17^{+0.09}_{-0.17})\pi$ \cr       
 $K_{31} $ & 1 &  ~~~~ $0-1.3$\cr
 $K_{32} $ & 1 &  ~~~~ $0.3-1.5$\cr
\hline
\end{tabular}
\end{center}
\end{table}

%%%%%%%%%%%%%%%%%%%%%%%%%%%%%%%%%%%%%%%%%%%%%%

%%%%%%%%%%%%%%%%%%%%%%%%%%%%%%%%%%%%%%%%%%%%%%%%%%%%%%%%%%%
\begin{figure}[t]
  \begin{center}
\includegraphics[width=10cm]{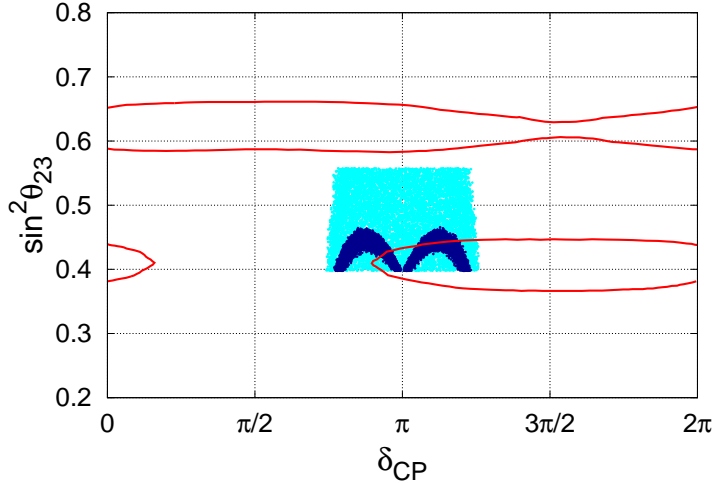}
    \caption{Predicted $\delta_{CP}$ versus $\sin ^2\theta _{23}$. 
 Dark blue (thick) region denotes the prediction when $K_{31}=K_{32}=1$,
      and, cyan (thin) shows the prediction for $K_{31}\not =1$ and/or
      $K_{32}\not= 1$.  All predictions are at 2$\sigma$.  The regions
      inside red contours are allowed in NO$\nu$A experiment at
      $1\sigma$ \cite{NOVA}.}
  \end{center}
\end{figure}
%%%%%%%%%%%%%%%%%%%%%%%%%%%%%%%%%%%%%%%%%%%%%%%%%%%%%%%%%%%%

We extend the model lifting the {\it ad~hoc} assumption that $K_{31}$
and $K_{32}$ are equal to unity (see ref.\cite{Obara:2006ny}). 
%%%%%%%%%%%%%%%%%%%%
%%%%%%%%%%%%%%%%%%%%
We keep the reality of the matrix.
We give in column 3 of Table 1, the neutrino mass parameters in this
extended model.  We find a limit on $K_{31}$ that it must be smaller
than 1.3, else we are led to too small a $\sin \theta_{23}$ to be
compatible with experiment.  There is no lower limit for $K_{31}$.  We
see that $m_1\propto K_{31}$, in so far as $m_1\ll (\Delta m^2_{12})^{1/2}$, so $m_1$ can vanish. 
 We find that mixing angles vary 
little towards the limit $K_{31}\rightarrow 0$, i.e., the agreement is
kept with experiment: only $m_1$ becomes small.
%In practice, we quit our calculations at $K_{31}=10^{-4}$.
We also find the limit $0.3<K_{32}<1.5$, the upper limit from the
lower limit of $\theta_{23}$, and the lower limit from $\theta_{13}$
to keep it not too small compared to experiment.

It may be appropriate to comment that the simple relation between
mixing angles and mass ratios is lost in the extended model, when $K_{31}$
is far from unity, as one can easily see for a two generation example.
Namely, $m_1=0$ does not mean a vanishing relevant mixing angle.  In
fact, in our case we see that the mixing angles change only a little,
as we take a limit $K_{31}\rightarrow 0$.

When we allow $K_{31}<1$, $ m_{ee}$ may be smaller: for
$K_{31}<0.2$, $m_{ee}$ takes $3-3.8$ meV.  On the other hand, $m_{ee}$
can be as large as 9 meV for $K_{31}\approx 1/2$ and $K_{32}\approx
1/2$. This is the upper limit of $m_{ee}$ attainable in our model.  Here, $m_1$
gives a dominant contribution to $m_{ee}$.

Figure 1 shows our prediction of $\delta_{CP}$ versus
$\sin^2\theta_{23}$ at 2$\sigma$
\footnote{One may calculate the
  two Majorana phases in the phase matrix
  $P={diag} \{1, \exp{(i \alpha/2)}, \exp{(i\beta/2)}\} $,
if $U_\nu$ would be cast into $U_\nu=VP$, from matrix elements
  together with $\sigma$ and $\tau$ we obtained. For our original FTY model,
  thus obtained $\alpha$ is small but non-zero ($\sim\pm13^\circ$).
  In the extended case $K\ne0$ the solution includes $\alpha=0$.
  $\beta$ is poorly determined, including zero. We do not discuss
  these phases further, as they do not directly appear in experiment. 
}.
This CP phase is correlated significantly with  $\sin^2\theta_{23}$  than with other mixing angles.  
Thick symbols stand for our
original $K_{31}=K_{32}=1$ model, and thin symbols show the region
allowed for $K_{31} \ne 1$ and $K_{32} \ne 1$.  We also draw contours
obtained by current NO$\nu$A experiment at 1$\sigma$.  A significant
overlap is still seen between the prediction and experiment
\cite{NOVA}.  T2K experiment reported earlier favours a finite CP
violating phase \cite{T2K0}.  Their result is consistent with the newly
reported NO$\nu$A experiment. It
is not shown in the figure, however, because their analysis assumes a
fixed value, $\sin^2\theta_{23}=0.5$.
%because $\sin^2\theta_{23}$ dependence of $\delta_{CP}$ is not
%available at present.

The extended model allows $\sin ^2\theta_{23}$ from 0.4 to
0.55, including maximal mixing 0.5.  We see that the region allowed
for $\delta_{CP}$ extends only little upon the inclusion of free
$K_{31}$  and $K_{32}$ parameters.
%A better determination of this phase in the future would give us
%a decisive test of the minimal texture model:
If a more accurate experiment in the future would fall in the region
outside the prediction, it compels that we must introduce a phase in
the neutrino mass matrix. This serves as a decisive test of the
current minimal texture model with neutrino mass matrix elements being real.

%%%%%%%%%%%%%%%%%%%%%%%%%%%%%%%%%%%%%%%%%%%%%%
%%%  Addition %%%%%%%%%%%%%%%%%%%%%%%%%%%%%%%%
%%%  Addition %%%%%%%%%%%%%%%%%%%%%%%%%%%%%%%%
%%%  Addition %%%%%%%%%%%%%%%%%%%%%%%%%%%%%%%%
%%%%%%%%%%%%%%%%%%%%%%%%%%%%%%%%%%%%%%%%%%%%%%
%{\bf 
%We can also predict the Majorana phases, $\alpha$ and $\beta$, which are given
%in the diagonal phase matrix ${diag} \{1, \exp{(i \alpha/2)}, \exp{(i\beta/2)}\%}$.  

What is relevant to leptogenesis is the factor $[(m_{\nu D} m_{\nu
    D}^\dagger)_{i3}]^2$ to which the phase contributes.
%in which, in fact, the phase of $B_\nu$ in
%the third row of eq.(1) appears in the (1,3) matrix element of
%$[m_{\nu D} m_{\nu D}^\dagger]$ being with $[(m_{\nu D} m_{\nu
%    D}^\dagger)_{12}]=0$.  
Whether lepton or antilepton excess is
determined by sign of the phase of $[(m_{\nu D} m_{\nu D}^\dagger)_{i3}]^2$,
%and that of $M_1-M_2$ and of $M_2-M_3$. 
and that of $M_i-M_3$ .
 In our analysis, e.g., both
$K_{31}>1$ and $<1$ are still allowed.
%, and so are $K_{32}$.
%%%%%%%%%%%%%%%%%%%%%%%%%%%%%%%%%%%%%%%%%%%%%%%%%
%%%  Addition %%%%%%%%%%%%%%%%%%%%%%%%%%%%%%%%
%%%  Addition %%%%%%%%%%%%%%%%%%%%%%%%%%%%%%%%
%%%%%%%%%%%%%%%%%%%%%%%%%%%%%%%%%%%%%%%%%%%%%
{The relation between the amount of lepton asymmetry and those
  phases, however, depends on details of the model beyond the mixing
  matrix, most
  importantly on the relative right-handed neutrino masses (i.e., the
  size of $K_{ij}$), which we cannot constrain from experiment at
  the current accuracy and our current knowledge. Therefore, we do not
  discuss leptogenesis further in the present paper.}
%%%%%%%%%%%%%%%%%%%%%%%%%%%%%%%%%%%%%%%%%%%%%%%%%%

 When the oscillation data becomes more accurate, we may hope that the sign of
the mass difference may eventually be predicted within the model
without resorting to a new type of the experimental information.

While the agreement of the current model with recent precision
experiment does not preclude the presence of the phase in the neutrino
mass matrix, the disagreement, on the other hand, would compel us, in
so far as we keep minimal texture, to
introduce a phase.  This is the phase that is needed to cause lepton
asymmetry, or leptogenesis.

%%%%%%%%%%%%%%%%%%%%%%%%%%%%%%%%%%%%%%%%%%%%%%%%%%%%%%%%%%%%%%

We conclude that our minimal texture model with real neutrino mass
matrix, devised when neutrino oscillation was first reported, passed
all tests concerning neutrino mixing that have been newly raised over
two decades; the predictions progressively improved upon new
experiments from time to time turned out to be so far all consistent
with later experiments.  It describes accurately the neutrino mixing
parameters available today, and as a result it is now tightly
constrained so that we can predict the CP violating phase. We proposed
here, as the final test for the minimal neutrino mass matrix, a test
whether predicted $\delta_{CP}$ agrees with experiment.

If it does, there is no compelling reason to introduce complex matrix
elements in the neutrino mass matrix, i.e., no compelling reason that
leptogenesis should occur in this model. 
%\cite{Ahn:2007mj}. 
If experiment would give
$\delta_{CP}$ deviated from our prediction, we are led to have intrinsically
complex matrix element that means a phase that is responsible for
leptogenesis. Neutrinoless double beta decay, if it would give 
$m_{ee}>9$ meV, also falsifies our minimal texture model.

Our analysis gives an example that the phase to be found in neutrino
oscillation does not mean the phase that causes leptogenesis. This
phase can be detected when real neutrino mass matrix fails to describe
the phase derived from neutrino oscillation.  More generally, this is
an example of the strategy as to how to detect the phase intrinsic to
the neutrino, requiring first the neutrino mass matrix to be strictly
real and detecting if any deviation from it needed.

%%%%% acknowledgement %%%%%
\vspace{1 cm}
\noindent
{\bf Acknowledgement}

This work is supported by JSPS Grants-in-Aid for Scientific Research
Nos. 15K05019 (MF), 16J05332 (YS),
15K05045, 16H00862 (MT) and 26287039, 26104009, 16H02176 (TTY).
M.F. is
supported by the Monell Foundation in Princeton.  This work receives a
support at IPMU by World Premier International Research Center
Initiative of the Ministry of Education in Japan.

\end{document}